\documentclass[10pt,a4paper]{article}
\pdfoutput=1
\usepackage{amsmath}
\usepackage{epstopdf}
\usepackage{hyperref}
\usepackage{graphicx,lscape}
\usepackage{footmisc}

\title{\Large \bf
Earthquakes economic costs through rank-size laws\thanks{We are
grateful to the Editor and three anonymous reviewers for valuable
comments. We also acknowledge the fruitful discussions with Marco
Cattaneo, Carlo Doglioni and Anna Maria Lombardi. All the remaining
errors are solely our responsibility.}}
\author{Valerio Ficcadenti, Roy Cerqueti\thanks{Corresponding author}}
\date{
University of Macerata, Department of Economics and Law,\\  via
Crescimbeni 20, I-62100, Macerata, Italy \\  $e$-$mail$ $addresses$:
v.ficcadenti@unimc.it (V. Ficcadenti), roy.cerqueti@unimc.it (R.
Cerqueti).}

\begin{document}
 \maketitle

\begin{abstract}
This paper is devoted to assess the presence of some regularities in
the magnitudes of the earthquakes in Italy between January
$24^{th}$, 2016 and January $24^{th}$, 2017, and to propose an
earthquakes cost indicator. The considered data includes the
catastrophic events in Amatrice and in Marche region. To our
purpose, we implement two typologies of rank-size analysis: the
classical Zipf-Mandelbrot law and the so-called universal law
proposed by Cerqueti and Ausloos (2016). The proposed generic
measure of the economic impact of earthquakes moves from the
assumption of the existence of a cause-effect relation between
earthquakes magnitudes and economic costs. At this aim, we
hypothesize that such a relation can be formalized in a functional
way to show how infrastructure resistance affects the cost. Results
allow us to clarify the impact of an earthquake on the social
context and might serve for strengthen the struggle against the
dramatic outcomes of such natural phenomena.
\end{abstract}

\section{Introduction}\label{Introduction}

Seismologists have carefully clustered the world in different
non-overlapping zones on the basis of the probability that the zone
experiences an earthquake. Such natural phenomena might cause very
dramatic damages to the human activities and kill several people.
Thus, policymakers should adopt anti-seismic building strategies,
mainly in zones with a high seismic risk. Unfortunately, some
countries come from a political history of myopic decisions in this
respect, and Italy is an illustrative example of them.
\newline
This paper aims at exploring the Italian earthquakes occurred in
2016 and early 2017, with a specific reference to the big ones in
Amatrice (August, $24^{th}$) and Visso (October, $26^{th}$ -- two
times -- and $30^{th}$) along with the large amount of minor
earthquakes before and after them. The considered period is 365
days, from January 24, 2016 to January 24, 2017, along which we
observe 978 seismic events within a Richter magnitude range: [3.1 -
6.5].  We decide to exclude observations with magnitudes smaller
than 3.1 for many reasons. First of all, this paper deals with
formulations of damages' cost indicators of the earthquakes and
according to the United States Geological Survey, a seismic event
with magnitude less than 3.1 has very low probability to cause
observable damages. Secondly, the restriction to magnitudes not
smaller than 3.1 allows to face the incomplete catalog problem.
Indeed, we are analyzing a peculiar time period from a seismic point
of view. Such a period has given a lot of work to the Italian
National Institute of Geophysics and Vulcanology (INGV) because of
the high number of earthquakes concentrated in very short time and
of the intensity of them. In fact, after the mainshock of Amatrice,
SISMIKO, the coordinating body of the emergency seismic network at
INGV, was activated to install a temporary seismic network
integrated with the existing permanent network in the epicentral
area, but the risk that many aftershocks were not registered or not
revised remains high (see Moretti et al., 2016). On this point, some
scholars are actively working on the estimation of the catalog
completeness. For example, Marchetti et al. (2016) have estimated
$M_c = 2.7$ for the revised catalog of the seismic events occurred
immediately after the Amatrice's earthquake. In accord to Marchetti
et al's work, $M_c$ could rise to a maximum level of 3.1 (on this
topic see also Chiaraluce et al., 2017).
\newline
Moreover, our dataset has no particular peaks apart from those
showed in Figure \ref{TimeSeries} after August $24^{th}$. Then, from
the 24/01/2016 to 23/08/2016, we can consider $M_c = 2.5$, in accord
to Romashkova and Peresan (2013) and Schorlemmer et al. (2010).
\newline
Thus, the considered restriction to magnitudes greater than 3.1 let
prudentially the catalog incompleteness problem be quite negligible
in the reference period without affecting the cost analysis of the
earthquakes.
\newline
We propose here a rank-size approach for analyzing the earthquakes'
magnitudes sequence just described in order to assess the presence
of data regularities.
\newline
The rank-size relationship has been explored for several sets of
data and it is still at the center of the scientific debate. At its
inception, power law and Pareto distribution with unitary
coefficient, introduced in Zipf (1935, 1949) and denoted from there
as \textit{Zipf law}, has been suitably employed to provide a best
fit of the rank-size connections in the field of linguistics.
\newline
After the first applications, several contributions supporting the
validity of the Zipf law have appeared in the literature. In this
respect, we just mention some recent important papers: Ioannides and
Overman (2003), Gabaix and Ioannides (2004), Dimitrova and Ausloos
(2015), Cerqueti and Ausloos (2015) in the context of economic
geography; Montemurro (2001) and Piantadosi (2014) in linguistic;
Axtell (2001), Fujiwara (2004), Bottazzi at al (2015) in the
business size field; Li and Yan (2002) in biology; Levene, Borges
and Loizou (2001) and Maillart et al (2008) in informatics; Manaris
et al (2005) and Zanette (2006), in the context of music; Huang et
al (2008) in the context of fraud detection; Blasius and T\"{o}njes
(2009) in the gaming field. For a wide review of rank-size analysis
see Pinto et al (2012). However, some cases of rank-size
relationships fail to be well-fitted by Zipf law (see e.g. Rosen and
Resnick, 1980; Peng, 2010; Ioannides and Skouras, 2013; Matlaba et
al., 2013). By one side, such examples support the acknowledged lack
of a theoretical ground for this statistical regularity (see Fujita
et al., 1999; Fujita and Thisse, 2000); by the other side, they
represent a further hint for proceeding with the methodological
research, and construct more general laws.
\newline
Indeed, under a pure methodological point of view, several
extensions of the Zipf law have been introduced. The most prominent
examples are the Zipf-Mandelbrot law (ZML, hereafter; see
Mandelbrot, 1953, 1961; Fairthorne, 2005) and the Lavalette law (LL
hereafter; see Lavalette, 1966), which have been proven to provide a
spectacular fit of rank-size relations, even when Zipf law fails to
do it (see e.g. Cerqueti and Ausloos, 2015).
\newline
In this paper, we implement two general rank-size procedures: the
above-mentioned ZML and a \textit{universal law} (UL from now on),
which is an extension of the LL to a five parameters rule that has
been recently introduced by Cerqueti and Ausloos (2016). All fits
have been carried out through a Levenberg-Marquardt algorithm
(Levenberg 1944, Marquardt 1963, Lourakis  2005) with a restriction
on the parameters that have to be positive.
\newline
Furthermore, we have also discussed the economic costs of the
earthquakes. At this aim, we propose a new generic cost indicator
based on a suitable transformation of magnitudes into costs. As we
will see, such an indicator moves from the best fit procedures
implemented in the rank-size analysis phase, and it might be
effectively used for finalizing policies for the management of
seismic risks. We show how the cost indicator can be computed in the
special case of the analyzed earthquakes.
\newline
Rank-size relations have been introduced for the explanation of
seismological data and for the earthquakes magnitudes (see e.g.
Jaume, 2000; Wu, 2000;  Mega et al, 2003; Newman, 2005; Saichev and
Sornette, 2006; Pinto et al, 2012; Aguilar-San Juan and
Guzman-Vargas, 2013). However, this is the first paper which treats
very recent Italian seismic events under this perspective. Moreover,
to the best of our knowledge, there are no contributions in the
literature on the construction of a cost indicator for earthquakes
based on the rank-size laws.
\newline
In order to validate the obtained results, extra investigations on
two different datasets have been performed. The first deals with a
more global analysis on the basis of a suitable enlargement of the
dtaset. At this aim, we notice that an important change of Italian
seismic network is occurred in $16^{th}$ April, 2005, when the new
network for seismic events collection has been activated. From that
date the data elaboration system has sensibly increased and, in
order to deal with the incompleteness catalog problem, the accepted
average $M_c$ has been set to 2.5 (see Romashkova and Peresan 2013,
and Schorlemmer et al. 2010). Therefore, we have performed the
rank-size analysis on the data from the INGV catalog in the period
ranging from 16/04/2005 to 31/03/2017, with the restriction to
magnitudes not smaller than 2.5.
\newline
The second extra investigation is developed to face the effects of
space variables. In this case, the considered dataset has been
created by selecting the earthquakes with epicenters in the eight
adjacent provinces involved in the seismic sequence started with the
Amatrice's earthquake: Macerata, Perugia, Rieti, Ascoli Piceno,
L'Aquila, Teramo, Terni and Fermo (and respective coasts), from
24/01/2016 to 24/01/2017. In so doing, we are in line with
geophysicists who claim that taking a small region and a short time
period let the space effects be not relevant (see e.g. De Natale et
al., 1988). It is interesting to note that, as we will see, the
local analysis is not too different from the original one in terms
of the cardinality of the dataset, in that the most part of the
earthquakes in the reference period in Italy has occurred in such
eight provinces.
\newline
The rest of the paper is organized as follows: Section \ref{data} is
devoted to the description of the data and of the methodologies used
for performing the analysis. This section illustrates also the
procedure adopted for the identification of the earthquakes costs
and for the development of the cost indicator. Section
\ref{robustness} investigates the robustness of the reached results
by presenting the study of the global and local datasets. Section
\ref{results} proposes the results of the analysis, along with a
critical discussion of them. Last Section concludes and offers
directions for future research.

\section{Data and methodology}
\label{data}

This section is devoted to the description of the data on the
magnitudes of the earthquakes occurred in Italy in 2016 and early
2017. Furthermore, it contains the illustration of the
methodological tools used for analysis.

\subsection{Data}

\begin{center}
\begin{figure}
    \centering
    \includegraphics[scale = 0.8]{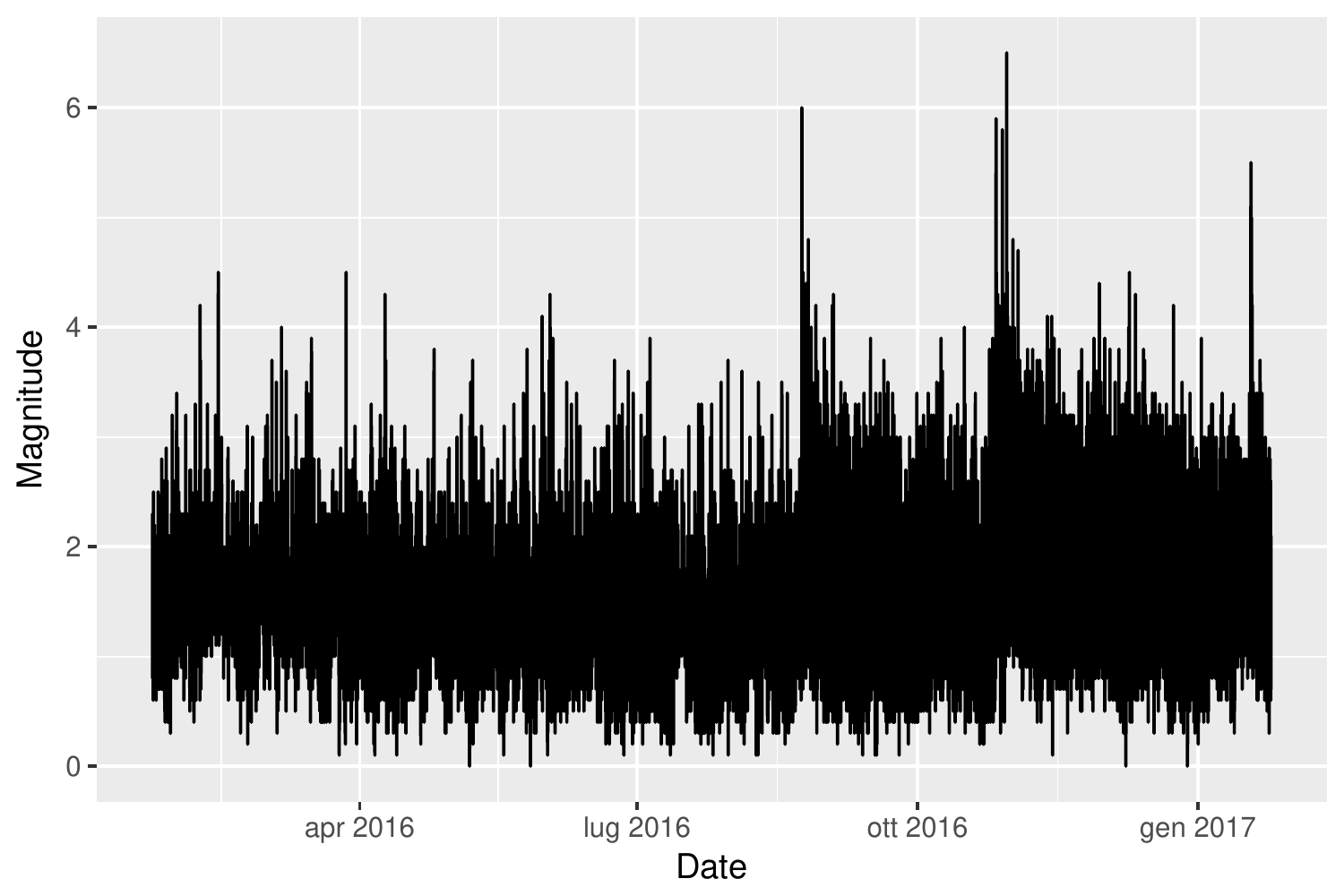}\\
    \caption{The time series of the earthquakes occurred from $24^{th}$ January 2016 to $24^{th}$ January 2017, according to the INGV data.
    The number of observations is 59190.}
    \label{TimeSeries}
\end{figure}
\end{center}

\begin{center}
\begin{figure}
    \centering
    \includegraphics[scale = 0.8]{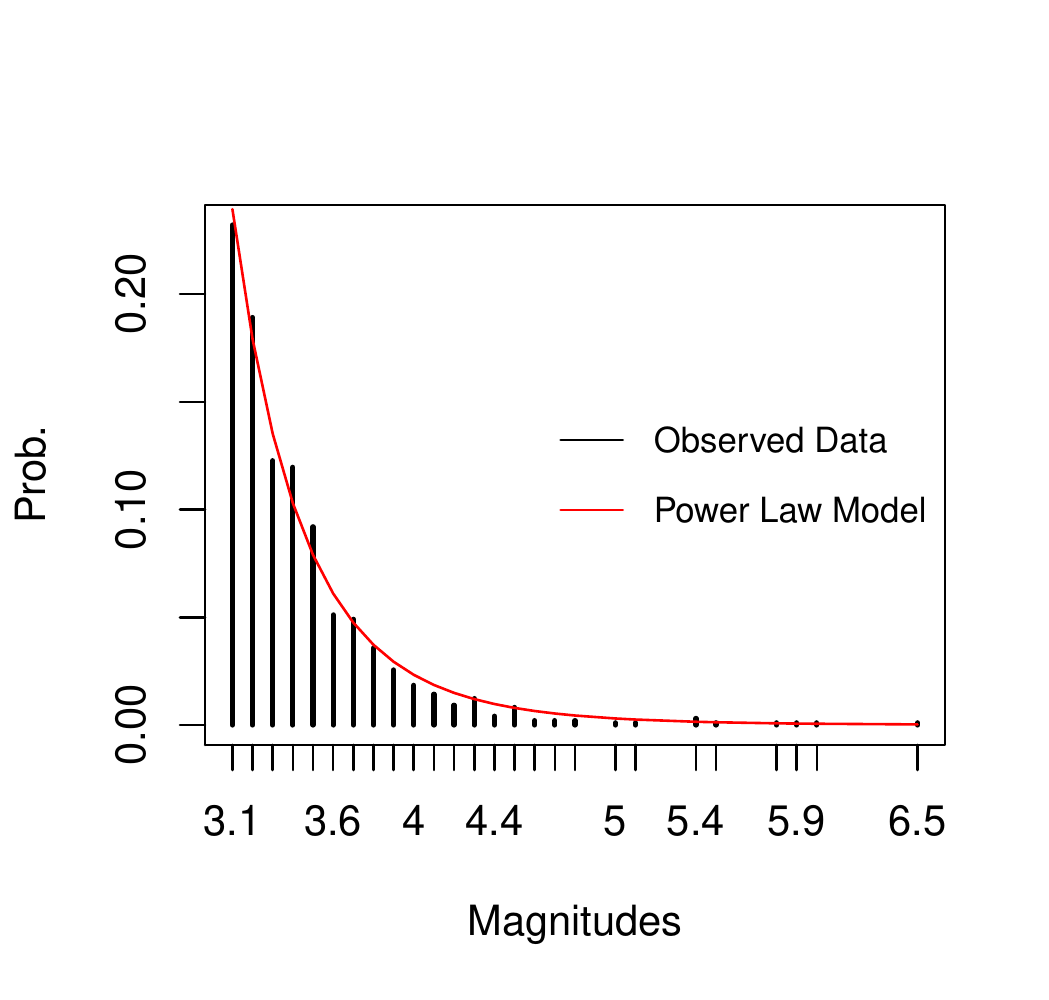}\\
\caption{Probability density function of all the earthquakes
registered from 24/01/2016 to 24/01/2017 with magnitudes not smaller
than 3.1. The best fit of the empirical distribution with a power
law of the type $y=ax^b$ is also shown. The calibrated parameters
are $\hat{a}=7428.58$ and $\hat{b}=-9.14$, with an $R^2$ of 0.99.}
    \label{PDF_data_3.1}
\end{figure}
\end{center}

Our dataset is composed by the magnitudes of the earthquakes
registered in Italy during the period: January $24^{th}$, 2016 -
January $24^{th}$, 2017.
\newline
The definition of the magnitude of an earthquake and the employed
dataset are taken from the website of the INGV (the Italian National
Institute of Geophysics and Vulcanology see
\url{<http://cnt.rm.ingv.it/>}). Such a definition is based on the
different measurement methods used from seismograms, each of them
being also tailored on a specific magnitude range and epicentral
distance. For the details on the concept of magnitude, please refer
to the website of the INGV (see
\url{<http://cnt.rm.ingv.it/en/help/>}).
\newline
Specifically, the considered period starts at the first hour of
January $24^{th}$,2016 and ends on the midnight of January
$24^{th}$, 2017, hence including relevant earthquakes like those
registered in Amatrice, on August $24^{th}$ (magnitude equals to 6)
Umbria and Marche regions on October $26^{th}$ (two times) and
$30^{th}$ of 2016 (magnitudes 5.4, 5.9 and 6.5 respectively), and
the most recent on January $18^{th}$ 2017, in L'Aquila (three times,
magnitude 5.5, 5.4 and 5.1). To have an idea of the seismic activity
of the analyzed period, see Figure \ref{TimeSeries}.
The number of the available data is of high relevance. Indeed, the
number of registered seismic events over the considered period is
59190, which gives to the reader the dimension of how often
earthquakes are registered in this period in Italy, in particular in
the Center of Italy, since the majority of the earthquakes are
located there. Data on depth of the epicenters and on their
localization are also available, but they are not treated in this
study. They are left for future researches.\\
We need to point out that there is a catalog incompleteness problem,
in that the main events might hide several minor subsequent
aftershocks. In order to deal with such catalog incompleteness
problem, we restrict the analysis to the seismic events of magnitude
not smaller than 3.1 (see Section \ref{Introduction} for a detailed
discussion of this point). Therefore, the number of observations
reduces to 978. Table \ref{Table1} collects the main statistical
indicators of the data and Figure \ref{PDF_data_3.1} represent the
probability density function of the considered time series. Notice
that Figure \ref{PDF_data_3.1} contains also the best fit of a power
law function with the empirical distribution of the sizes of the
earthquakes. This supports an empirical evidence, already pointed
out by previous studies (see e.g. Kagan, 2010). Some comments on the
statistical characteristics can be found in Section \ref{results}.

\begin{table} \begin{center}
\begin{tabular}[t]{cc}
  \hline
  \textbf{Statistical indicator} & \textbf{Value} \\
\hline
Number of data  & 978\\
Maximum  & 6.50\\
Minimum & 3.10\\
Mean ($\mu$) & 3.42\\
Median ($m$) & 3.30\\
RMS & 3.45\\
Standard Deviation ($\sigma$) & 0.39\\
Variance & 0.15\\
Standard Error & 0.01\\
Skewness & 2.67\\
Kurtosis & 14.36\\  \hline
 $\mu/\sigma$ & 8.73\\
$3(\mu-m)/\sigma$ & 0.95 \\
\hline
\end{tabular}
   \caption{Summary of the statistical
characteristics for the magnitudes not smaller than 3.1 of the
earthquakes in Italy during 365 days: from January $24^{th}$, 2016
to  January $24^{th}$, 2017.} \label{Table1}
\end{center}
\end{table}

\subsection{Methodologies}
\label{method}
The magnitude of an earthquake represents the size of the rank-size
analysis.
\newline
Since the target of the analysis is to construct an aggregated costs
indicator, magnitudes are not taken as they are. Indeed, the same
earthquake can produce different levels of damages if it follows a
long list of foreshocks or not: in the former case, the earthquake
insists over an already solicited territory, while in the latter one
it is the first shake and human activities have not previous
solicitations. Therefore, each earthquake has been temporally
contextualized -- suppose, it has occurred at time $t$ -- and we
have transformed its magnitude $z$ into $\tilde{z}=\eta(n,z_1,\dots,
z_n,\Delta t)\times z$, where $\eta(n,z_1,\dots, z_n,\Delta t)$ is a
parameter dependent on the number $n$ of the foreshocks whose
magnitudes are assumed to be $z_1,\dots, z_n$ and occurred in the
time interval $[t-\Delta t,t]$. The parameter $\eta(n,z_1,\dots,
z_n,\Delta t)$ is marginally increasing with respect to $z_1, \dots,
z_n$ and $n$ and marginally decreasing with respect to $\Delta t$,
and it is not smaller than 1. In fact, if the territory has
experienced several foreshocks of large magnitude in a small time
range before $t$, then the damages created by the earthquake are
comparable with those of an isolated earthquake with magnitude
$\tilde{z}>z$. \newline With a reasonable abuse of notation, we
refer hereafter simply to magnitudes, having in mind $\tilde{z}$
instead of $z$.
\newline
The single earthquakes have been ranked in decreasing order, so that
rank $r=1$ corresponds to the highest registered magnitude while
$r=978$ is associated to the lowest value of the considered
phenomenon, which is 3.1. Then, in general, low ranks are the ones
associated to the strongest seismic events in terms of magnitudes,
while high ranks point to the earthquakes with small magnitudes.
\newline
Here we implement two times the best fit procedure to assess
whenever the size-magnitude $z$ might be view as a function of the
rank $r$. The considered fit functions are the ZML and the UL. The
former can be written as
\begin{equation}
\label{ZML} \tilde{z} \sim f_{ZML}(r)=\alpha(r+\beta)^{-\gamma},
\end{equation}
while the latter is
\begin{equation}
 \tilde{z} \sim f_{UL}(r)=k\dfrac{\left(N+1-r+\psi
\right)^{\xi}}{\left[N(r+\phi)\right]^\lambda}, \label{UL}
\end{equation}
where $\alpha$, $\beta$, $\gamma$ must be calibrated on the size
data when (\ref{ZML}) is used, while $k$, $\psi$, $\xi$, $\phi$,
$\lambda$ are those to be calibrated if the fit procedure is as in
(\ref{UL}). The parameter $N$ corresponds to the number of
observations, and it is $N=978$ for this specific case.
\newline
To implement the rank-size analysis and derive the proposed
aggregated cost indicator we need to provide an explicit shape of
the parameter $\eta(n,z_1,\dots, z_n,\Delta t)$. In order to meet
space constraints\footnote{The proposal of other scenarios and their
analysis is available upon request.}, we present here the analysis
of the unbiased scenario of $\eta(n,z_1,\dots, z_n,\Delta t)=1$, for
each $n,z_1,\dots, z_n,\Delta t$. In this case we are in absence of
amplification effects. Since we aim at constructing an aggregated
cost indicator, this situation has an intuitive reasoning: indeed,
it is the case with the lowest level of damages -- all the
earthquakes are treated as isolated ones -- and let clearly
understand how the outcomes of a missing anti-seismic policy can be
negative, even in the lucky case of absence of propagation effects.
\newline
Under the considered scenario, we have $\tilde{z}=z$.
\newline
The economic indicator is obtained by transforming the magnitude of
an earthquake into the cost associated to such an earthquake. In
this respect, as already said above, the decision of taking
magnitudes not smaller than 3.1 lies also in the evidence that a
very low-magnitude earthquake does not produce damages. We assume
that costs are positive and increasing for magnitudes greater than a
certain threshold $\bar{z} \geq 3.1$, and they are null below it.
The value of the critical threshold $\bar{z}$ is strongly affected
by the way in which infrastructures and buildings are constructed on
the seismic territory. Neglecting the adoption of anti-seismic
building procedures leads to destructive earthquakes even at low
magnitudes, i.e. when $\bar{z}$ has a small value.
\newline
Under a general perspective, we use the rank-size laws written in
(\ref{ZML}) and (\ref{UL}) in order to transform magnitudes into
costs. This will lead to the definition of two different cost
indicators, as we will see.
\newline
We define $C_\star:[0,+\infty) \rightarrow[0,+\infty)$ such that
$C_\star(z)=H(f_\star(r))$, where $\star=ZML, UL$. Quantity
$C_\star(z)$ is the cost associated to an earthquake with magnitude
$z$ when the best fit is performed through function $f_\star$ and
$H:[0,+\infty) \rightarrow[0,+\infty) $ increases in $[\bar{z},
+\infty)$ and is null in $[0,\bar{z})$.
\newline
Under the rank-size law perspective, the identification of a
critical magnitude $\bar{z}$ is associated to the identification of
a critical rank $\bar{r}$ such that $z \leq \bar{z}$ if and only if
$r \geq \bar{r}$. Such a critical rank varies if one takes
(\ref{ZML}) and (\ref{UL}). To distinguish them, we will refer to
the intuitive notation of $\bar{r}_{ZML}$ and $\bar{r}_{UL}$.
\newline
The cost indicator $\Gamma$ associated to the collection of the
considered earthquakes is defined as the aggregation of their
individual costs. We include in such an aggregation also the
presence of a maximum for the level of magnitude of an earthquake,
and we denote it by $Z_{MAX}$. In fact, we point out that the
greatest magnitude ever registered is 9.5 of the Great Chilean
earthquake in 1960. To be prudential, we will set a theoretical
$Z_{MAX}=10$ even if the empirical maximum is 6.5, as reported in
the applications (see Table \ref{Table1}).
\newline
Thus, we set
\begin{equation}
    \label{GammaZML}
    \Gamma_{ZML} = \int_{\bar{z}}^{Z_{MAX}}C_{ZML}(z)dz = \int_{0}^{\bar{r}_{ZML}}H\left(\hat{\alpha}(r+\hat{\beta})^{-\hat{\gamma}}\right) dr,
\end{equation}
and
\begin{equation}
    \label{GammaUL}
    \Gamma_{UL} = \int_{\bar{z}}^{Z_{MAX}}C_{UL}(z)dz =
    \int_{0}^{\bar{r}_{ZML}}H\left(\hat{k}\dfrac{(N+1-r+\hat{\psi})^{\hat{\xi}}}{[N(r+\hat{\phi})]^{\hat{\lambda}}}\right) dr,
\end{equation}
which represent the cost indicators for the fits in (\ref{ZML}) and
(\ref{UL}), respectively, and where $\hat{\star}$ is the calibrated
parameter $\star$, according to the best fit procedure.
\newline
The $\Gamma$'s depend on the value of $\bar{z}$, once all the rest
is fixed. Of course, the cost indicators decrease as $\bar{z}$
increases, and they are null when $\bar{z}=Z_{MAX}$.
\newline
We propose three scenarios for the selection of function $H$:
\begin{itemize}
 \item[$(i)$] \[H(z)= \begin{cases}
    \exp(z),& \forall z \in [\bar{z}, Z_{MAX}];\\
    0,       & \forall z \in [0,\bar{z});
\end{cases}
\]

\item[$(ii)$] \[H(z)= \begin{cases}
    z,& \forall z \in [\bar{z}, Z_{MAX}];\\
    0,       & \forall z \in [0,\bar{z});
\end{cases}
\]
\item[$(iii)$] \[H(z)= \begin{cases}
    \ln(z),& \forall z \in [\bar{z}, Z_{MAX}];\\
    0,       & \forall z \in [0,\bar{z});
\end{cases}
\]
\end{itemize}
The considered scenarios are representative of three very different
realities for the economic costs. Indeed, the exponential case (item
$(i)$) is the one providing a severe penalization of the high
magnitudes in terms of costs; differently, the logarithm (item
$(iii)$) is the function assigning a lower value to the costs for
high magnitudes and the linear case (item $(ii)$) is the middle case
between these extremes.
\newline
To identify the considered cases, we will insert an intuitive
superscript to the cost indicator so that, for example,
$\Gamma^{(ii)}_{ZML}$ is the $\Gamma_{ZML}$ obtained when $H$ is as
in item $(ii)$.

\section{Robustness check}
\label{robustness}

\begin{center}
\begin{figure}
    \centering
    \includegraphics[scale = 0.8]{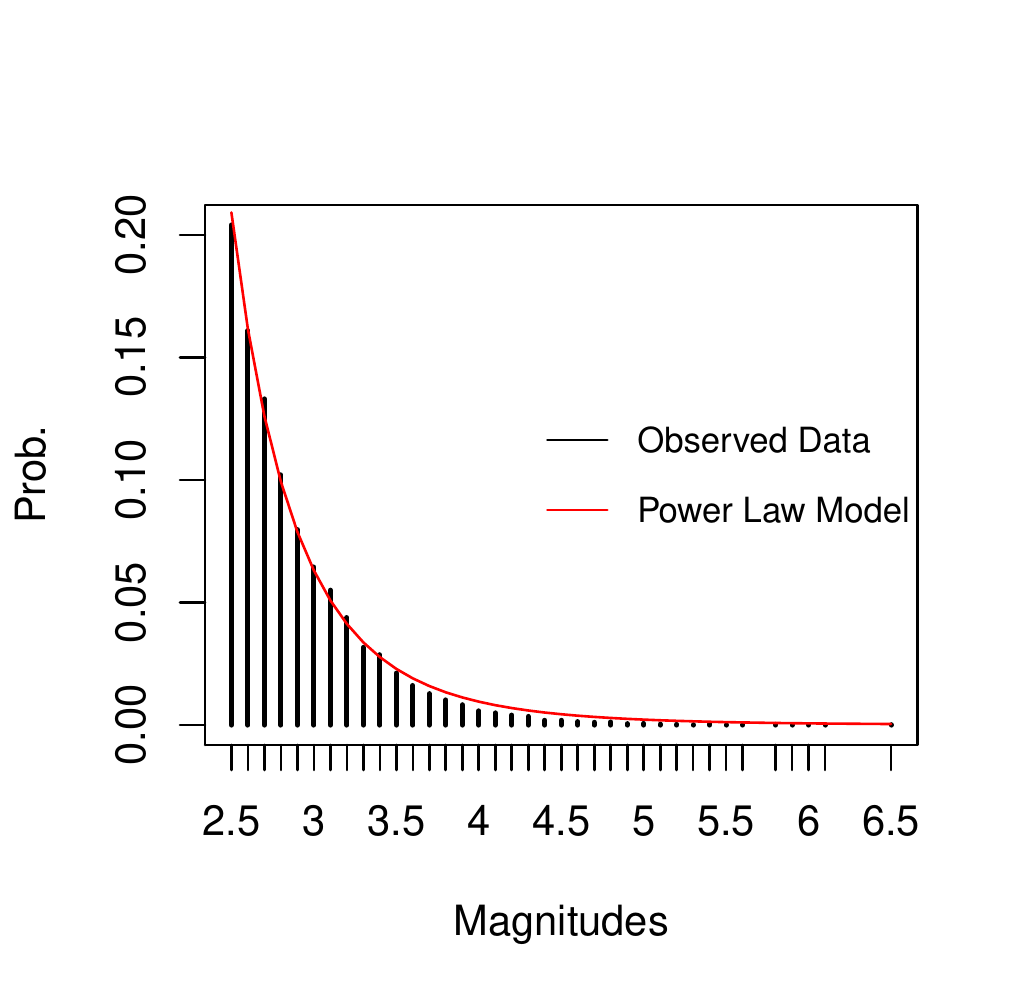}\\
    \caption{Probability density
function of all the earthquakes registered from 16/04/2005 to
31/03/2017 with magnitudes not smaller than 2.5. The best fit of the
empirical distribution with a power law of the type $y=ax^b$ is also
shown. The calibrated parameters are $\hat{a}=86.32$ and
$\hat{b}=-6.57$, with an $R^2$ of about 1.}
    \label{PDF_data_2.5}
\end{figure}
\end{center}

\begin{center}
\begin{figure}
    \centering
    \includegraphics[scale = 0.8]{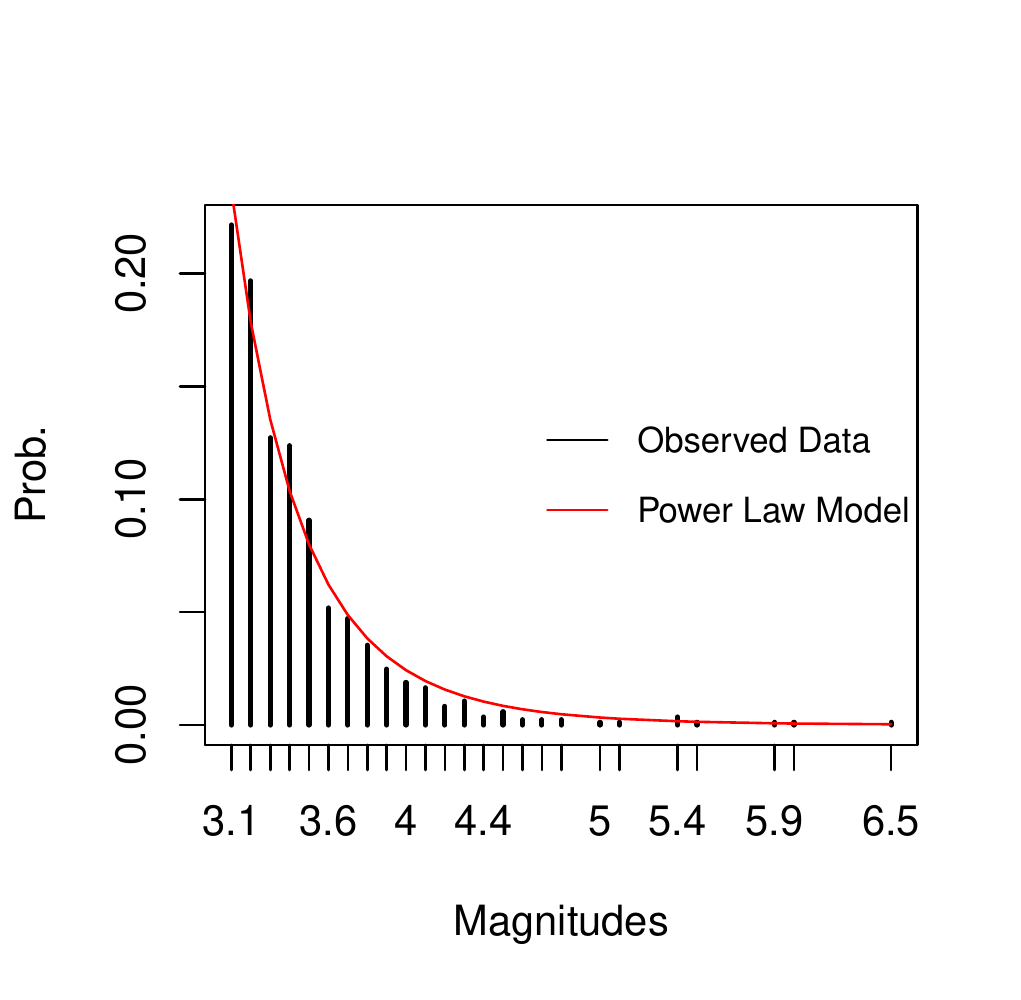}\\
    \caption{Probability density
function of all the earthquakes registered in the provinces of
Macerata, Perugia, Rieti, Ascoli Piceno, L'Aquila, Teramo, Terni and
Fermo from January $24^{th}$, 2016 to January $24^{th}$, 2017, with
magnitudes not smaller than 3.1. The best fit of the empirical
distribution with a power law of the type $y=ax^b$ is also shown.
The calibrated parameters are $\hat{a}=5805.79$ and $\hat{b}=-8.93$,
with an $R^2$ of 0.98.}
    \label{PDF_data_3.1_local}
\end{figure}
\end{center}

In order to validate the obtained findings, we here investigate the
problem by using two different datasets: a global and a local one.
\newline
In the global case, we present the analysis on a bigger dataset by
assuming that enlarging the considered time window let the average
magnitude completeness be closer to 2.5, in accord to Romashkova and
Peresan (2013), and Schorlemmer et al. (2010). In so doing, we
provide a validation of the results.  So, we have downloaded from
the same source (INGV), 13239 observations detected from April
$16^{th}$, 2005 to March $31^{st}$, 2017 with magnitude not smaller
than 2.5. The initial data is consistently selected, in that it
coincides with the change of the Italian earthquake survey by INGV.
Table \ref{summarydataenlargedwindows} contains a summary statistics
of the dataset and in Figure \ref{PDF_data_2.5} there is the
probability density function of the data. As for the original
sample, Figure \ref{PDF_data_2.5} shows that a power law is a good
approximation of the empirical distribution of the earthquakes (see
e.g. Kagan, 2010). Table \ref{bigtimewindowestimatations}
illustrates the parameters of the best fit estimation obtained by
applying the processes described in Section \ref{method} on this
global dataset. For a visual inspection of the estimated model,
refer to Figures \ref{ZML_fit_2.5_largedata} and
\ref{RS5_fit_2.5_largedata}, which contain the original data and the
fitted model of the calibration performed with Eq. (\ref{ZML}) and
(\ref{UL}) respectively.
\newline
In the local case, we explore the spatial effects by running the
same procedure described in Section \ref{method} on the restricted
area of the provinces of Macerata, Perugia, Rieti, Ascoli Piceno,
L'Aquila, Teramo, Terni and Fermo (for the estimation precision of
the epicenters see Amato and Mele, 2008) that are relevant for the
2016 Amatrice earthquake sequence (see Gruppo di Lavoro INGV sul
Terremoto in Centro Italia, 2016). The reference period is the same
of the original study: from January $24^{th}$, 2016 to January
$24^{th}$ 2017, with 849 observations. This local analysis is in
line, from a methodological point of view, with seismological
researches which state that taking small zones and short time
periods leads to negligible space effects (see e.g. De Natale et
al., 1988). Notice that the local analysis serves as validating the
robustness of the study of the considered sample. This said, it is
also important to stress that the identification of an earthquake as
a product of spatio-temporal correlations among shakes is not
relevant for implementing the rank-size analysis and, subsequently,
for deriving the aggregated cost indicator. Indeed, we are not
interested on the reasoning behind the occurrence of an earthquake
but only on the fact that it has occurred and on the knowledge of
its magnitude. To be sure that we avoid the catalog incompleteness
and in order to make the analysis comparable with the one object of
this paper, we take in consideration magnitudes not smaller than 3.1
(Marchetti et al., 2016). It is very important to note that the
local dataset contains about the $87\%$ of the earthquakes of the
original sample. Thus, results of the local analysis in line with
those obtained for the original sample are expected. The statistical
summary of the reduced dataset is reported in Table
\ref{summarydatalocal} while the density function of the registered
magnitudes is presented in Figure \ref{PDF_data_3.1_local}. Also in
this case, Figure \ref{PDF_data_3.1_local} evidences that the
empirical distribution of the earthquakes follows a power law (see
e.g. Kagan, 2010). Table \ref{localestimatations} contains the
parameters of the best fit estimation obtained by applying the
processes described in Section \ref{method} on the local data. For a
visual inspection of the estimated model, Figures
\ref{ZML_fit_3.1_localdata} and \ref{RS5_fit_3.1_localdata} contain
the original data and the fitted model of the calibration performed
with Eq. (\ref{ZML}) and (\ref{UL}) respectively.

\begin{table}[ht]
\centering
\begin{tabular}{cc}
  \hline
 \textbf{Statistical Indicator} & \textbf{Value} \\
  \hline
Number of Data & 13239 \\
  Maximum & 6.50 \\
  Minimum & 2.50 \\
  Mean ($\mu$)& 2.88 \\
  Median ($m$) & 2.80 \\
  RMS & 2.91 \\
  Standard Deviation ($\sigma$) & 0.42 \\
  Variance & 0.18 \\
  Standard Error & 0.002 \\
  Skewness & 1.89 \\
  Kurtosis & 8.24 \\ \hline
  $\mu/\sigma$ & 6.84 \\
  $3(\mu-m)/\sigma$ & 0.60 \\
   \hline
\end{tabular}
\caption{Summary of the statistical characteristics for the
magnitudes not smaller than 2.5 of the earthquakes occurred from
April $16^{th}$, 2005 to March $31^{st}$, 2017.}
\label{summarydataenlargedwindows}
\end{table}

\begin{table}[ht]
\centering
\begin{tabular}{cc}
  \hline
 \textbf{Statistical Indicator} & \textbf{Value} \\
  \hline
Number of Data & 849 \\
  Maximum & 6.50 \\
  Minimum & 3.10 \\
  Mean ($\mu$) &  3.42 \\
  Median ($m$) & 3.30 \\
  RMS & 3.44 \\
  Standard Deviation ($\sigma$) & 0.39 \\
  Variance & 0.15 \\
  Standard Error & 0.01      \\
  Skewness & 2.75 \\
  Kurtosis & 15.05 \\ \hline
  $\mu/\sigma$ & 8.79 \\
  $3(\mu-m)/\sigma$ & 0.95 \\
   \hline
\end{tabular}\\
 \caption{Summary of the statistical
characteristics for the magnitudes of the earthquakes with
epicenters in the provinces of Macerata, Perugia, Rieti, Ascoli
Piceno, L'Aquila, Teramo, Terni and Fermo from January $24^{th}$,
2016 to January $24^{th}$, 2017.} \label{summarydatalocal}
\end{table}

\section{Results and discussion}
\label{results}

\begin{center}
\begin{figure}
    \centering
    \includegraphics[scale = 0.8]{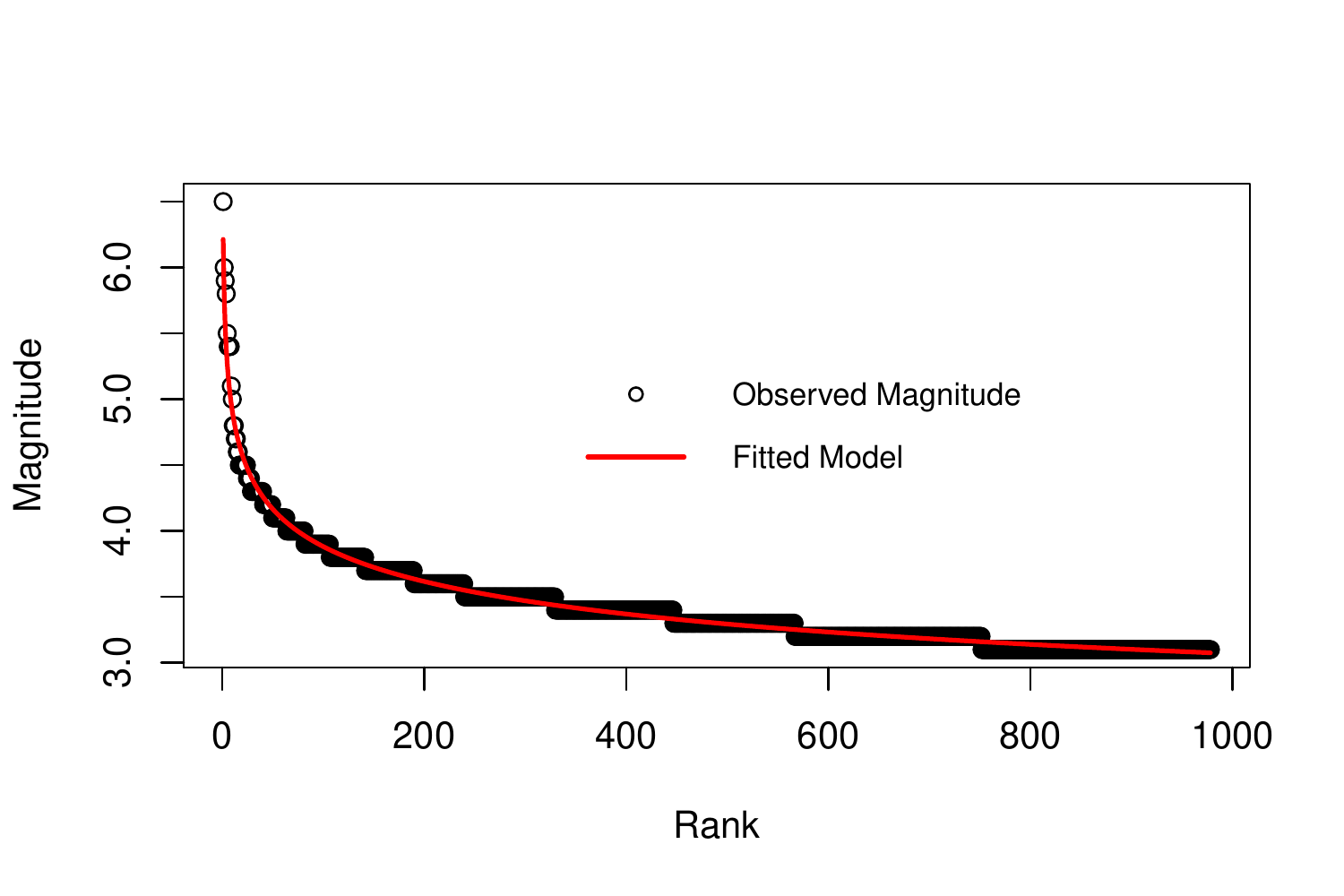}\\
\caption{All the earthquakes with magnitude not smaller than 3.1
registered in Italy from 24/01/2016 to 24/01/2017 ranked by
decreasing order according to their magnitude with the corresponding
ZML fit. See formula (\ref{ZML}).}
    \label{ZMLfit0}
\end{figure}
\end{center}

\begin{center}
\begin{figure}
    \centering
    \includegraphics[scale = 0.8]{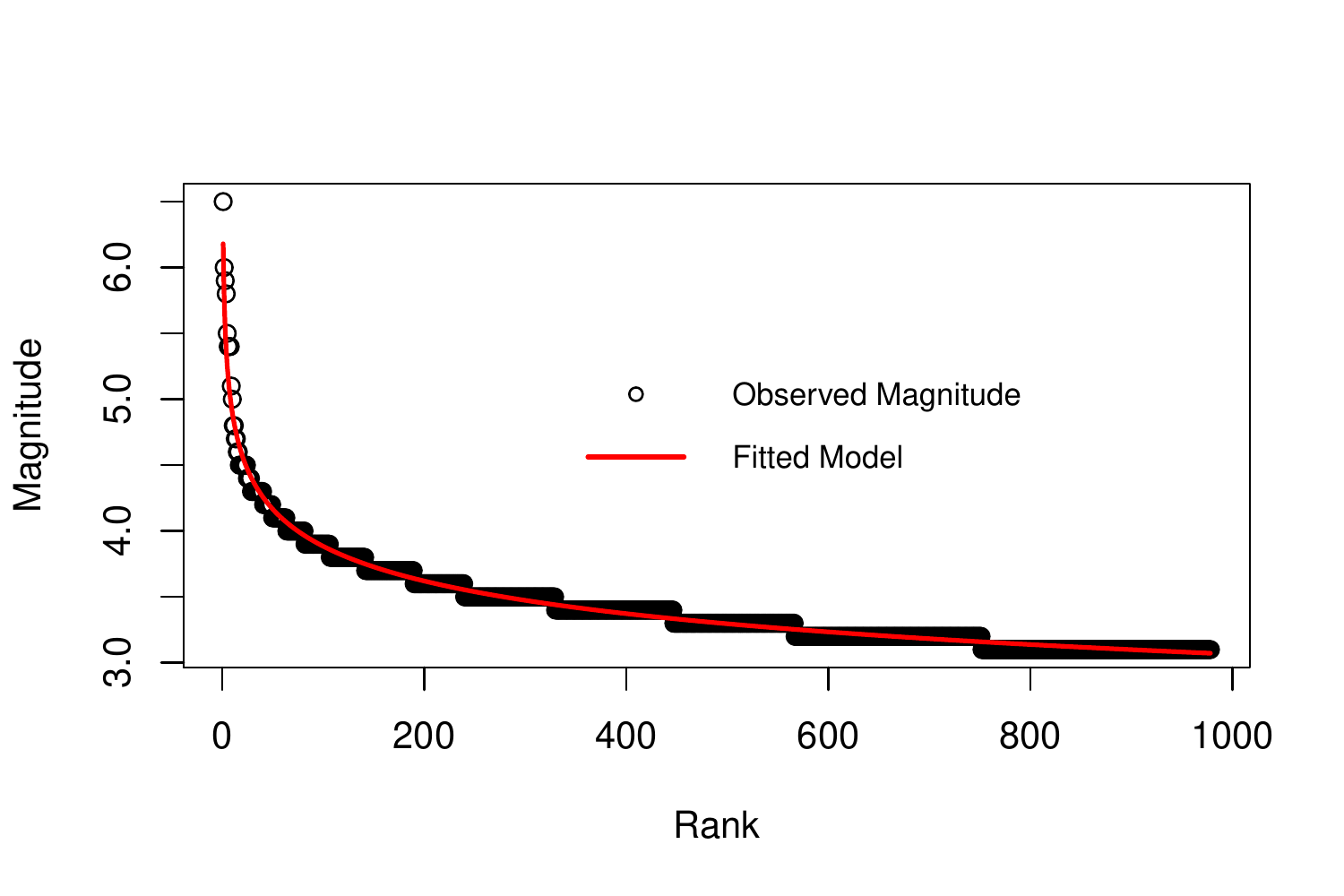}\\
\caption{All the earthquakes with magnitude not smaller than 3.1
registered in Italy from 24/01/2016 to 24/01/2017 ranked by
decreasing order according to their magnitude with the corresponding
UL fit. See formula (\ref{UL}).}
    \label{ULfit0}
\end{figure}
\end{center}

Table \ref{Table1} offers a preliminary view of the phenomenon under
investigation. Since the empirical distribution of the sizes of the
earthquakes can be well-fitted through a power law, as expected, the
mean and the median of the magnitude distribution are different.
This suggests the presence of asymmetry. The positional indicators
show that the most part of the observations takes values close to
3.3. Furthermore, the variability indexes confirm that the values
are rather concentrated near the distribution's center. The positive
skewness suggests a right-tailed shape, and the value of the
kurtosis indicates a leptokurtic distribution. The leptokurtic
property of the data is due to the presence of outliers (see Figure
\ref{PDF_data_3.1}).
\newline
As mentioned above, the best fit procedures on (\ref{ZML}) and
(\ref{UL}) are performed over the dataset considering magnitudes not
smaller than 3.1 for the reasons discussed in Section
\ref{Introduction} and Section \ref{robustness}. Results are
presented in Table \ref{Table2} where the calibrated parameters and
the $R^2$'s are reported. For a visual inspection of the goodness of
fit, refer to Figures \ref{ZMLfit0} and \ref{ULfit0}.

\begin{table}
\begin{center}
\begin{tabular}[t]{ccc}
  \hline \hline
  Eq. (\ref{ZML}) & \textbf{Calibrated parameter} & \textbf{Value}
  \\
\hline
&$\hat{\alpha}$& 6.21\\
& $\hat{\beta} $& 0.00\\
& $\hat{\gamma}$ & 0.10\\
\hline & $R^2$ & 0.98\\
\hline \hline
  Eq. (\ref{UL}) & \textbf{Calibrated parameter} & \textbf{Value} \\
  \hline
& $\hat{k}$& 8.63 \\
& $\hat{\phi} $& 0.00\\
& $\hat{\lambda}$& 0.10\\
& $\hat{\psi} $& 6972.72\\
& $\hat{\xi} $& 0.04\\
\hline & $R^2$ & 0.98\\
\hline \hline
\end{tabular}
   \caption{Calibrated parameters of the best fit procedures,
according to formulas (\ref{ZML}) and (\ref{UL}) for the dataset
with magnitude not smaller than 3.1 (N = 978; period: 24/01/2016 -
24/01/2017; Italy). The value of the $R^2$ in both of cases is
reported.}
   \label{Table2}
\end{center}
\end{table}

\begin{table}
\begin{center}
\begin{tabular}[t]{ccc}
  \hline \hline
  Eq. (\ref{ZML}) & \textbf{Calibrated parameter} & \textbf{Value}
  \\
\hline
&$\hat{\alpha}$& 9.48\\
& $\hat{\beta} $& 68.80\\
& $\hat{\gamma}$ & 0.14\\
\hline & $R^2$ & 0.98\\
\hline \hline
  Eq. (\ref{UL}) & \textbf{Calibrated parameter} & \textbf{Value} \\
  \hline
& $\hat{k}$& 0.88 \\
& $\hat{\phi} $& 9.52\\
& $\hat{\lambda}$& 0.11\\
& $\hat{\psi} $& 36951.95\\
& $\hat{\xi} $& 0.30\\
\hline & $R^2$ & 0.99\\
\hline \hline
\end{tabular}
   \caption{Calibrated parameters of the best fit procedures, according to formulas (\ref{ZML}) and (\ref{UL}) for the dataset
built on an enlarged time window: April $16^{th}$, 2005 - March
$31^{st}$, 2017 (N = 13239, magnitudes not smaller than 2.5). The
value of the $R^2$ in both of cases is reported.}
   \label{bigtimewindowestimatations}
\end{center}
\end{table}

\begin{center}
\begin{figure}
    \centering
    \includegraphics[scale = 0.8]{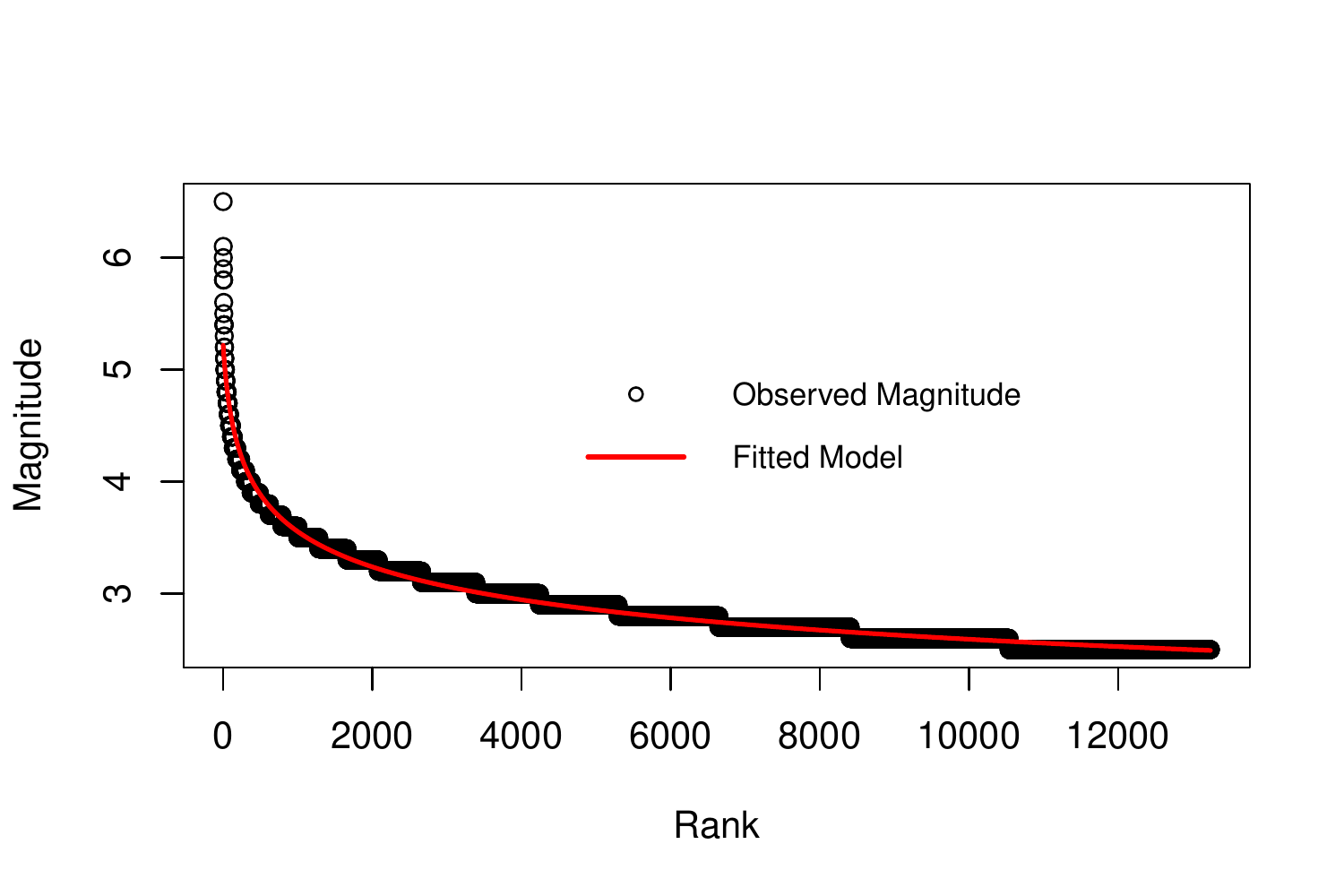}\\
    \caption{All the earthquakes registered from 16/04/2005 to 31/03/2017 with magnitudes not smaller than 2.5,
    ranked by decreasing order according to their magnitude with the corresponding ZML fit. See formula (\ref{ZML}).}
    \label{ZML_fit_2.5_largedata}
\end{figure}
\end{center}

\begin{center}
\begin{figure}
    \centering
    \includegraphics[scale = 0.8]{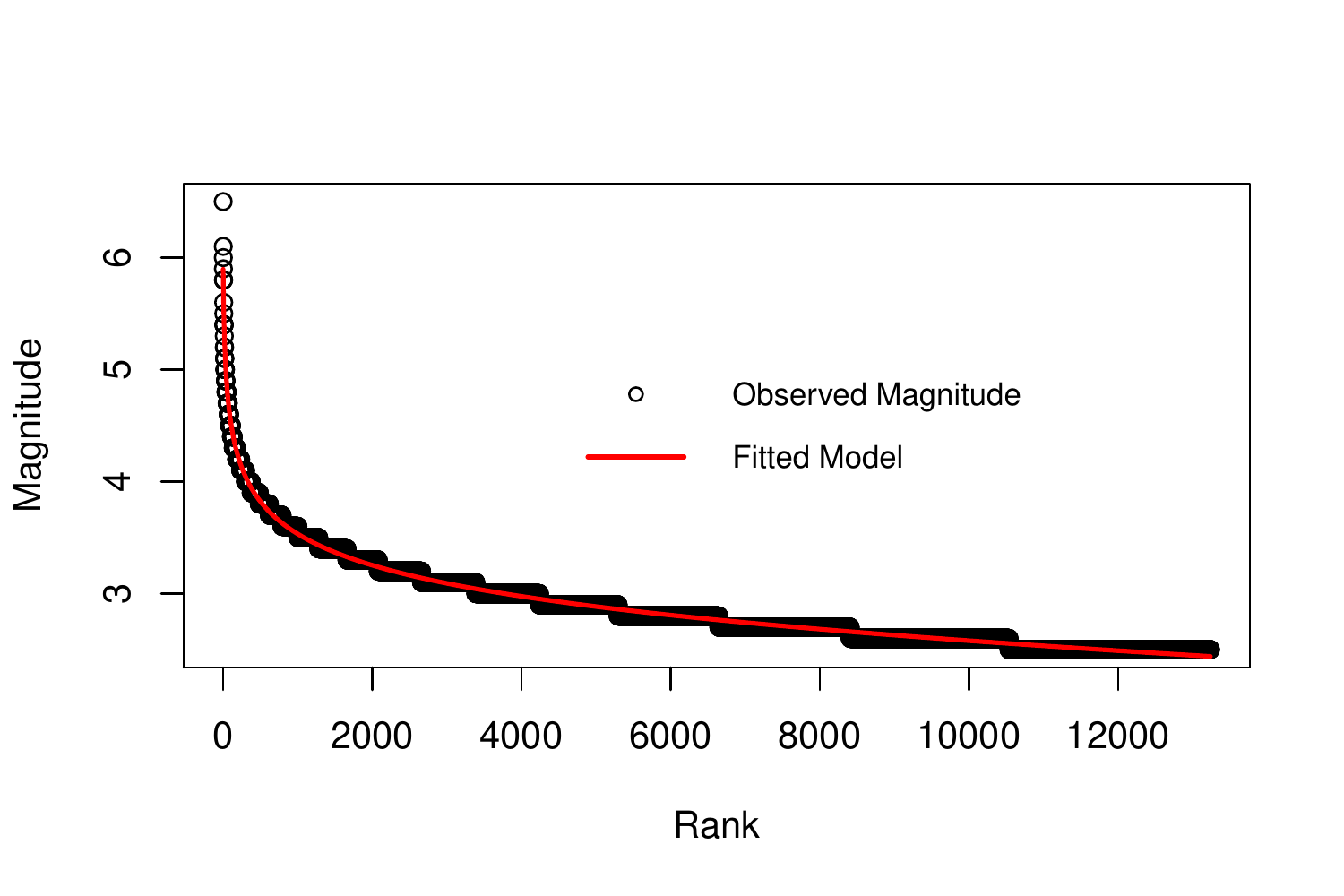}\\
    \caption{All the earthquakes registered from 16/04/2005 to 31/03/2017 with magnitudes not
    smaller than 2.5, ranked by decreasing order according to their magnitude with the corresponding UL fit. See formula (\ref{UL}).}
    \label{RS5_fit_2.5_largedata}
\end{figure}
\end{center}

\begin{table}
\begin{center}
\begin{tabular}[t]{ccc}
  \hline \hline
  Eq. (\ref{ZML}) & \textbf{Calibrated parameter} & \textbf{Value}
  \\
\hline
&$\hat{\alpha}$& 6.07\\
& $\hat{\beta} $& 0.00\\
& $\hat{\gamma}$ & 0.10\\
\hline & $R^2$ & 0.98\\
\hline \hline
  Eq. (\ref{UL}) & \textbf{Calibrated parameter} & \textbf{Value} \\
  \hline
& $\hat{k}$& 9.50 \\
& $\hat{\phi} $& 0.00\\
& $\hat{\lambda}$& 0.10\\
& $\hat{\psi} $& 6749.18\\
& $\hat{\xi} $& 0.02\\
\hline & $R^2$ & 0.98\\
\hline \hline
\end{tabular}
   \caption{Calibrated parameters of the
best fit procedures, according to formulas (\ref{ZML}) and
(\ref{UL}) for the dataset of the earthquakes occurred during the
period: January $24^{th}$, 2016 to January $24^{th}$, 2017, with
epicenters localized in the provinces of Macerata, Perugia, Rieti,
Ascoli Piceno, L'Aquila, Teramo, Terni and Fermo, N = 849,
magnitudes not smaller than 3.1. The value of the $R^2$ in both of
cases is reported.}
   \label{localestimatations}
\end{center}
\end{table}

\begin{center}
\begin{figure}
    \centering
    \includegraphics[scale = 0.8]{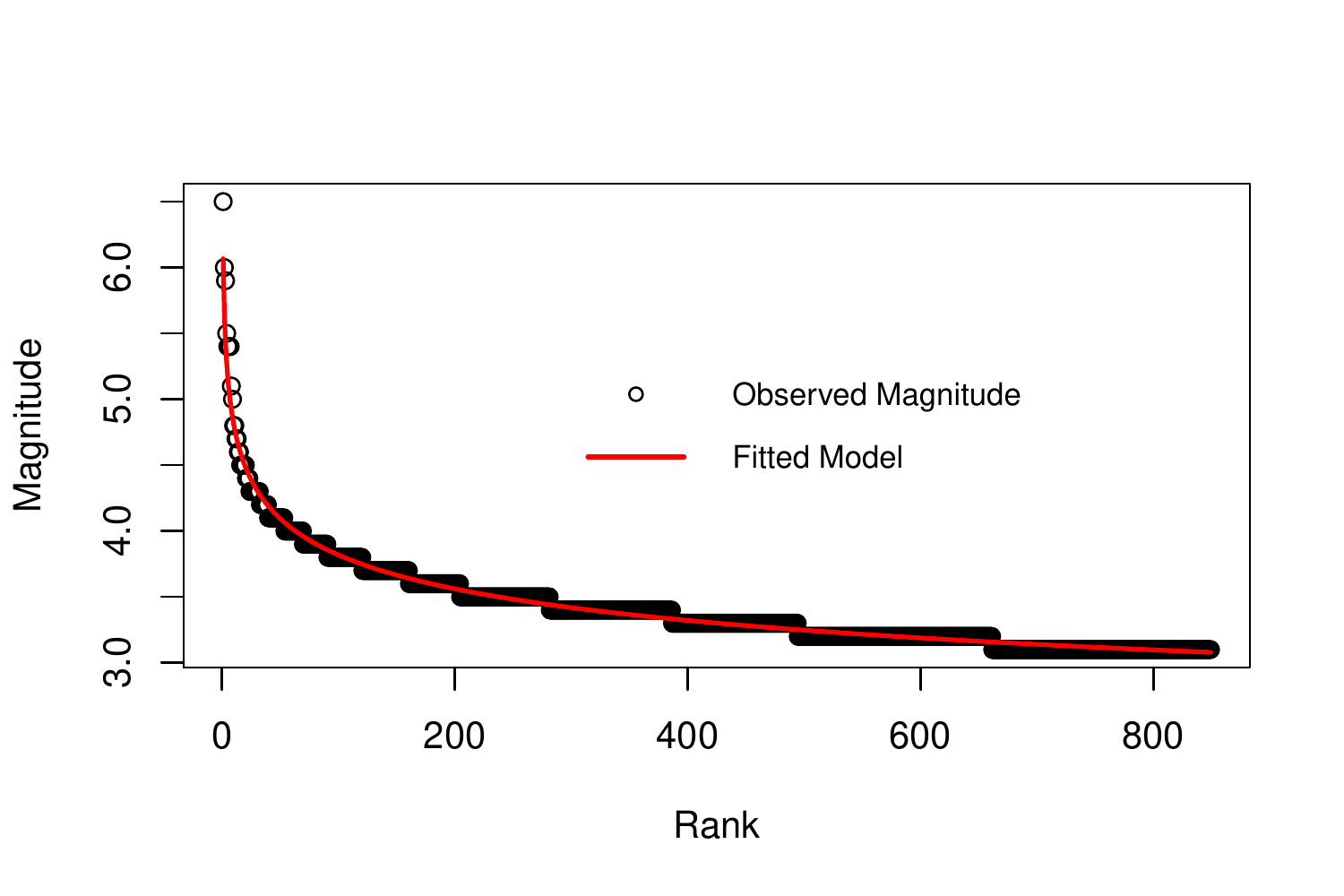}\\
    \caption{The earthquakes registered from 24/01/2016 to 24/01/2017 in the provinces of
Macerata, Perugia, Rieti, Ascoli Piceno, L'Aquila, Teramo, Terni and
Fermo, with magnitudes not smaller than 3.1, ranked by decreasing
order according to their magnitude with the corresponding ZML fit.
See formula (\ref{ZML}).}
    \label{ZML_fit_3.1_localdata}
\end{figure}
\end{center}

\begin{center}
\begin{figure}
    \centering
    \includegraphics[scale = 0.8]{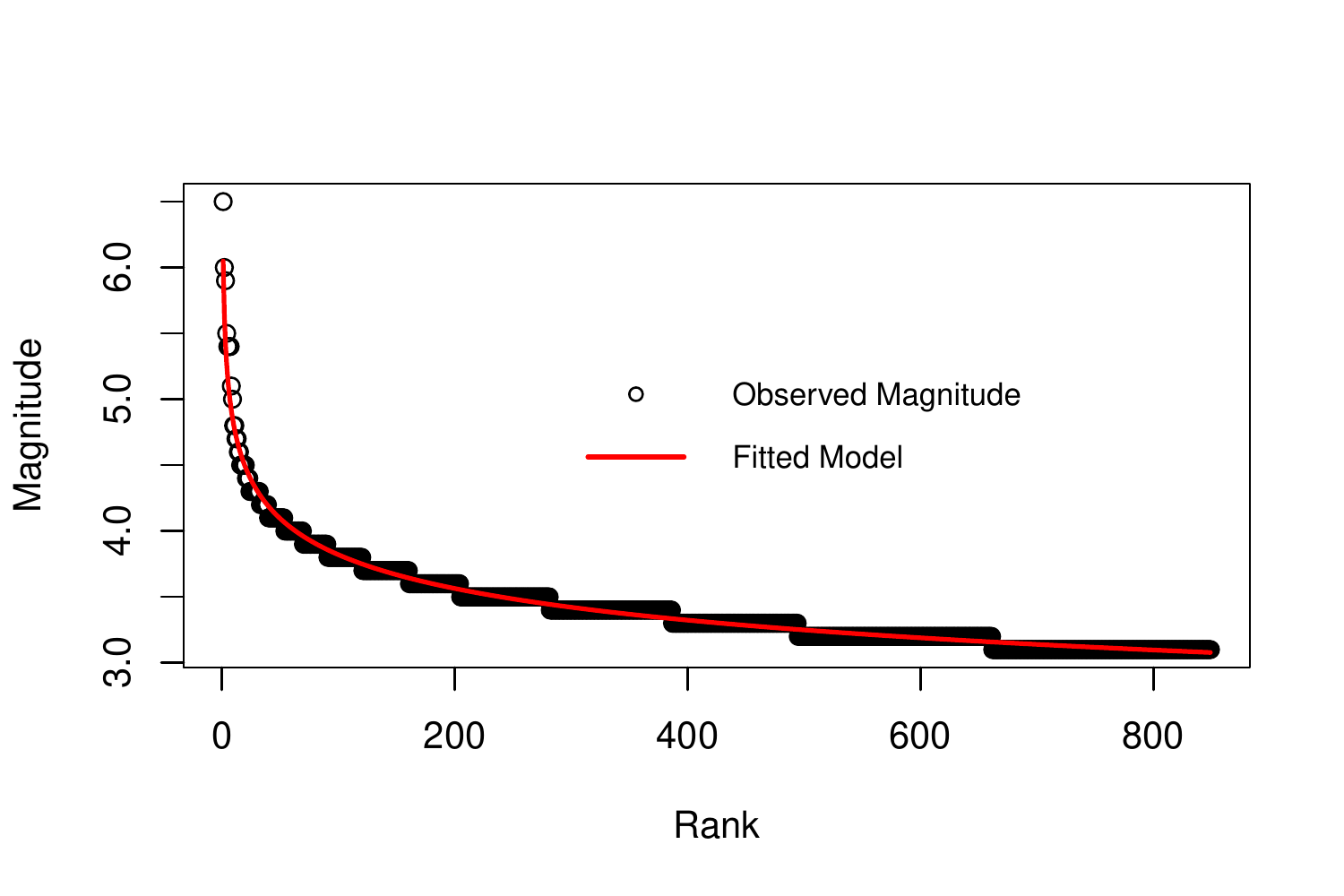}\\
    \caption{The earthquakes registered from 24/01/2016 to 24/01/2017
in the provinces of Macerata, Perugia, Rieti, Ascoli Piceno,
L'Aquila, Teramo, Terni and Fermo with magnitudes not smaller than
3.1, ranked by decreasing order according to their magnitude with
the corresponding UL fit. See formula (\ref{UL}).}
    \label{RS5_fit_3.1_localdata}
\end{figure}
\end{center}

The analysis evidences a first important fact that is the presence
of outliers at low ranks. They do not affect the performance of the
fitting procedures with
(\ref{ZML}) or (\ref{UL}), and consequently we cannot note substantial discrepancies in using ZML or UL for the dataset containing the earthquakes
from 24/01/2016 to 24/01/2017 in Italy.\\
Looking at Section \ref{robustness}, we can compare our results with
those obtained for the global and the local datasets and check the
coherence of our findings.
\newline
The local analysis excludes 149 observations with magnitudes mainly
allocated in the high rank and only one of magnitudes around 5. The
exclusions do not change too much the estimations, and the
parameters and the $R^2$'s remain rather similar to those presented
for the case of the original sample. Such a similarity appears to be
more evident for the ZML fit, hence supporting that the UL
approximates the data in a more convincing way and is more sensitive
to data variation (see Tables \ref{Table2} and
\ref{localestimatations}). In particular, the upper side of Tables
\ref{Table2} and \ref{localestimatations} shows a ZML best fit
calibration with $\hat{\beta}$ close to zero and a small value of
$\hat{\gamma}$ because the fitted model captures at the best the
effect of the low ranks. Consequently, $\hat{\alpha}$ is close to
the highest registered magnitude. Visual inspection is also
appealing (see Figures \ref{ZMLfit0} and \ref{ZML_fit_3.1_localdata}
for the ZML case and Figures \ref{ULfit0} and
\ref{RS5_fit_3.1_localdata} for the UL case). This suggests the
negligible presence of space effects in performing the rank-size
analysis and computing the cost indicators.
\newline
The situation is notably different for the case of the dataset with
enlarged time window (see Table \ref{bigtimewindowestimatations}).
In this case, we observe an increment of
the relative number of magnitudes at high ranks, hence leading to a
calibration which is more distorted from the small magnitude events
and loses representation capacity at lowest ranks, even in presence of some outliers at low ranks.\\
%
The opportunity to catch the effects of the lowest ranked outliers
is due to $\psi$ in (\ref{UL}) (see Cerqueti and Ausloos, 2016)
which increases in the case of sizes at low ranked magnitudes close
to the medium ranked sizes. By comparing the levels of the parameter
$\hat{\psi}$ from Tables  \ref{Table2}, \ref{localestimatations} and \ref{bigtimewindowestimatations}, one can observe the
increment in the global case. Notice that a small value of $\hat{\psi}$
stands for a fit which can capture the high ranked data effect
without flattering the part of the curve at a low rank. Moreover,
the parameter $\phi$ in (\ref{UL}) acts in the same way of $\psi$,
but to capture the effects of the lowest outliers. Thus, in presence
of high ranked outliers the value of $\phi$ increases. Consistently
with this idea, $\hat{\phi}$ is equal to 9.52 for the case of the
enlarged time window and it is null in the other cases.
\newline
A slight improvement of the goodness of fit is shown by the $R^2$ of
the enlarged case, even if it moves from 0.98 to 0.99. So, the
goodness of fit is generally so high that a discrepancy between
observed data and fit curves are not appreciable (see Figures
\ref{ZMLfit0}, \ref{ZML_fit_2.5_largedata} and
\ref{ZML_fit_3.1_localdata} for the ZML case and Figures
\ref{ULfit0}, \ref{RS5_fit_2.5_largedata} and
\ref{RS5_fit_3.1_localdata} for the UL case).
\newline
We also notice that the highest (lowest) level of the magnitudes
estimated through (\ref{ZML}) and (\ref{UL}), namely
$\hat{Z}^{ZML}_{Max}$ and $\hat{Z}^{UL}_{Max}$
($\hat{Z}^{ZML}_{Min}$ and $\hat{Z}^{UL}_{Min}$), respectively, adds
further arguments for supporting the goodness of fit. In fact, we
have found $\hat{Z}^{ZML}_{Max} = 6.21$, $\hat{Z}^{UL}_{Max} =
6.18$, $\hat{Z}^{ZML}_{Min} = 3.07$ and $\hat{Z}^{UL}_{Min} = 3.07$.
For the maximum points curves are slightly below the maximum
empirical observation of 6.5, while for minimum we have the same
value very close to 3.1, hence suggesting an analogous behavior at
the highest rank.
\newline
To sum up, we argue that the ZML and UL show similar behaviors in
fitting the original catalog and the one associated to the local
dataset, hence giving a substantial lack of space effects. The
analysis of catalog with $M_c=2.5$ and wider time windows highlights
that the UL fit is more appropriate to represent the data, even if
the goodness of fit remains unchanged. Thus, data show an analogous
regularity property in both of cases of short and long period, and
this suggests that results provided for the original sample are
robust to enlargement of the period. The incompleteness catalog
problem has been faced in both of cases by truncating to a low level
of magnitude, in accord to seismological literature.
\newline
For what concerns the economic costs indicators, some integrals can
be easily computed in closed form, while other ones will be
estimated. We have

\begin{equation}
\label{Gamma^iiZML}
    \Gamma^{(ii)}_{ZML} = \int_{0}^{\bar{r}_{ZML}}\hat{\alpha}(r+\hat{\beta})^{-\hat{\gamma}} dr = \frac{\hat{\alpha}}{1-\hat{\gamma}}\left[(\bar{r}_{ZML}+\hat{\beta})^{1-\hat{\gamma}}-\hat{\beta}^{1-\hat{\gamma}}\right]
\end{equation}

\begin{equation}
\begin{split}
    \label{Gamma^iiiZML}
    \Gamma^{(iii)}_{ZML}&=\int_0^{\bar{r}_{ZML}}\ln\left(\hat{\alpha}(r+\hat{\beta})^{-\hat{\gamma}}\right)dr=
    \ln\left(\hat{\alpha}\right)\cdot\bar{r}_{ZML}-\\
    &-\hat{\gamma}\cdot\left[(\bar{r}_{ZML}+\hat{\beta})\{\ln (\bar{r}_{ZML}+\hat{\beta})-1\}-\hat{\beta}\{\ln (\hat{\beta})-1\}\right];
\end{split}
\end{equation}

\begin{equation}
\begin{split}
    \label{Gamma^iiiUL}
    \Gamma^{(iii)}_{UL} &= \int_0^{\bar{r}_{UL}} \ln\left(\hat{k}\cdot \dfrac{(N+1-r+\hat{\psi})^{\hat{\xi}}}{[N(r+\hat{\phi})]^{\hat{\lambda}}}\right) dr =
    \ln\hat{k}\cdot{\bar{r}_{UL}}+\\ &+\hat{\xi}\left[-(N+1-\bar{r}_{UL}+\hat{\psi})\{\ln(N+1-\bar{r}_{UL}+\hat{\psi})-1\}
    +(N+1+\hat{\psi})\{\ln(N+1+\hat{\psi})-1\}\right] - \\
    &-\hat{\lambda}\cdot \left[\ln (N)\cdot \bar{r}_{UL}
    +(\bar{r}_{UL}+\hat{\phi})\{\ln(\bar{r}_{UL}+\hat{\phi})-1\}-\hat{\phi}\{\ln(\hat{\phi})-1\}\right].
\end{split}
\end{equation}

\begin{center}
\begin{figure}
    \centering
    \includegraphics[scale = 0.6]{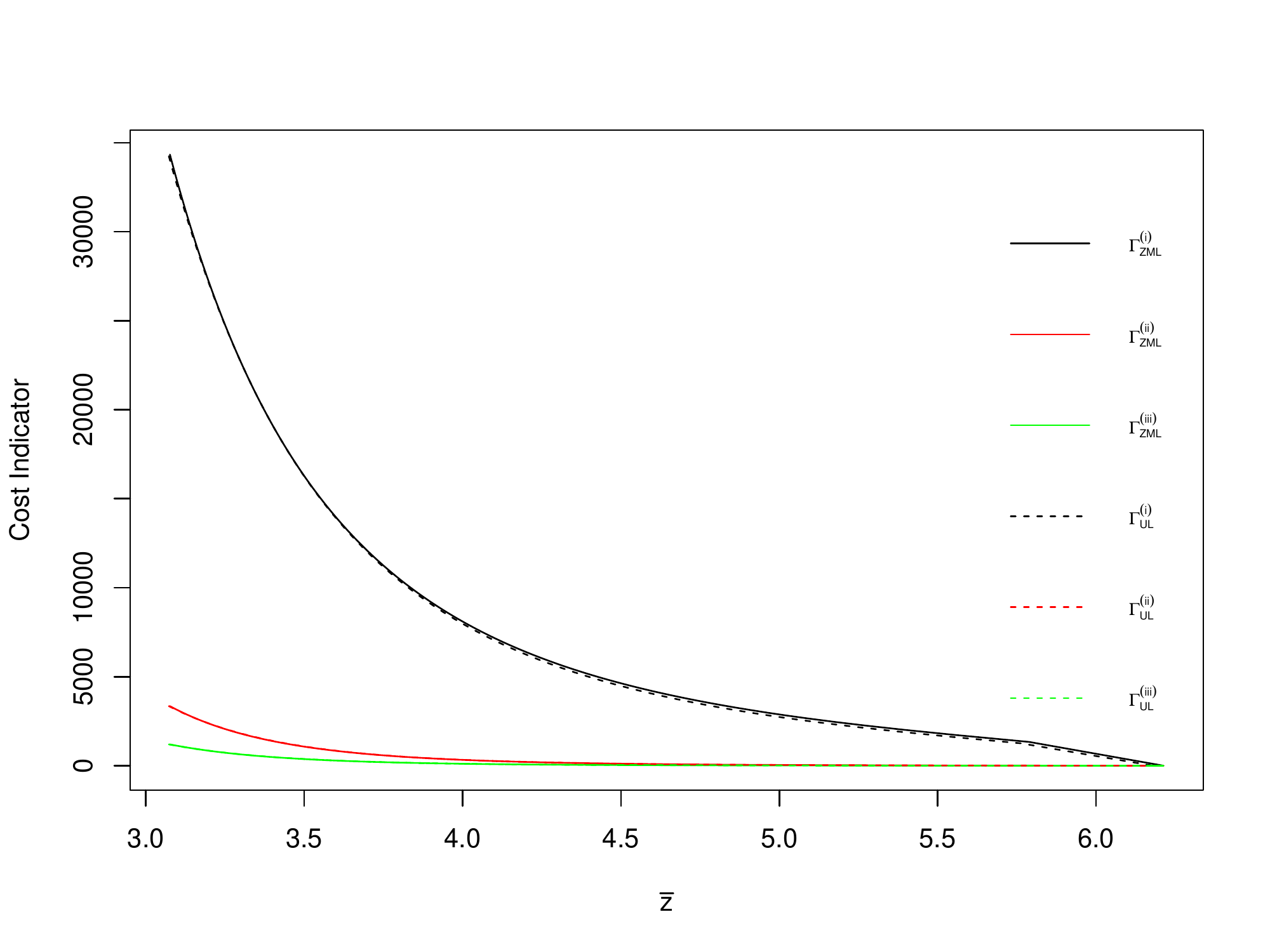}\\
    \caption{Comparison among $\Gamma^{(i)}_{ZML}$,
    $\Gamma^{(ii)}_{ZML}$, $\Gamma^{(iii)}_{ZML}$, $\Gamma^{(i)}_{UL}$, $\Gamma^{(ii)}_{UL}$ and $\Gamma^{(iii)}_{UL}$ as $\bar{z}$
    varies. They are calibrated on the Italian earthquakes registered from 24/01/2016 to 24/01/2017 with magnitudes not smaller than 3.1}
    \label{cost_analysis}
\end{figure}
\end{center}

\begin{center}
\begin{figure}
    \centering
    \includegraphics[scale = 0.6]{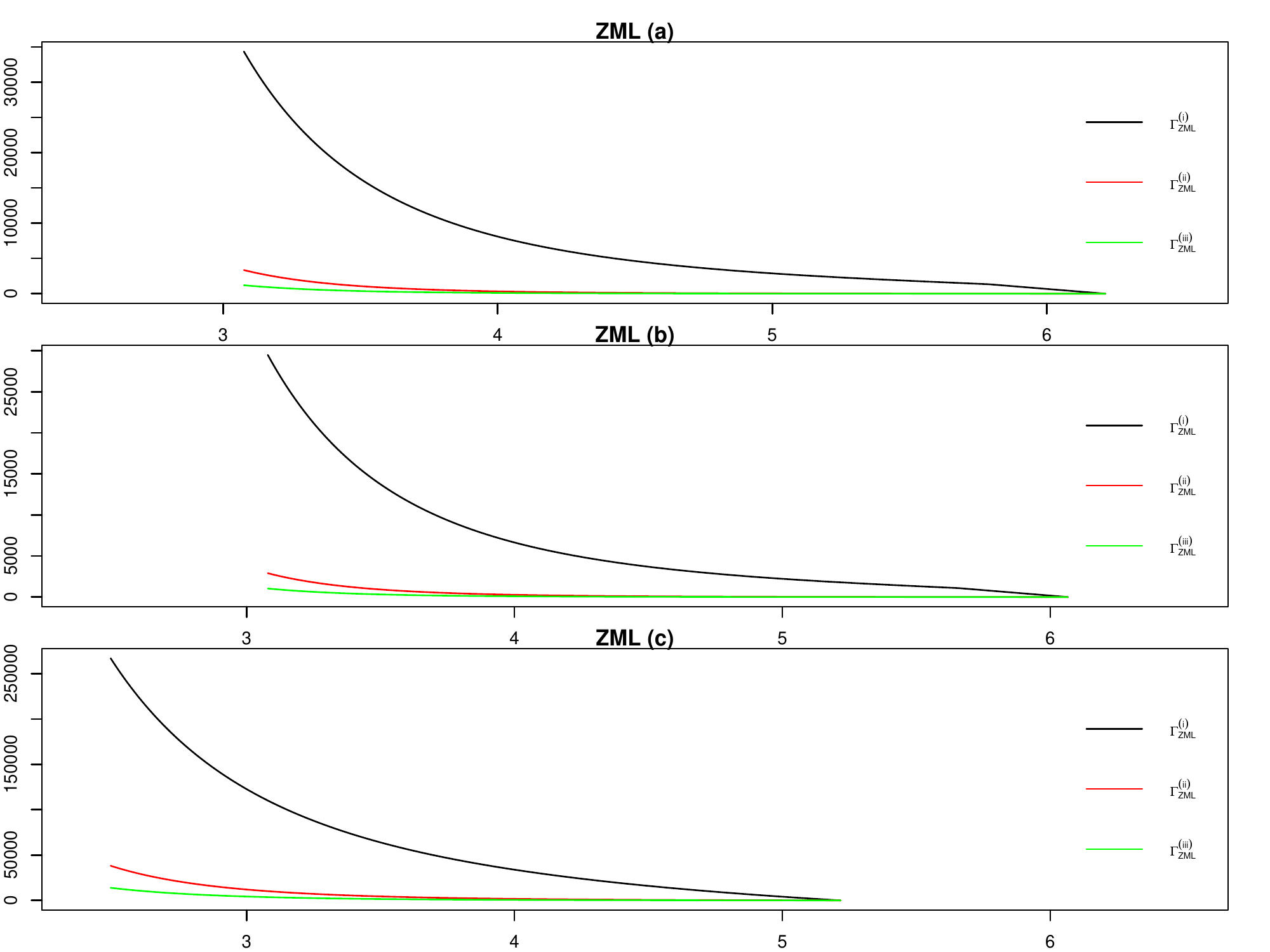}\\
    \caption{(a) Comparison among $\Gamma^{(i)}_{ZML}$,
    $\Gamma^{(ii)}_{ZML}$ and $\Gamma^{(iii)}_{ZML}$ as $\bar{z}$ varies. The case of earthquakes
    registered from 24/01/2016 to 24/01/2017 in Italy with magnitudes not smaller than 3.1 is presented. \newline
    (b) Comparison among $\Gamma^{(i)}_{ZML}$,
    $\Gamma^{(ii)}_{ZML}$ and $\Gamma^{(iii)}_{ZML}$ as $\bar{z}$ varies. The case of earthquakes
    registered from 24/01/2016 to 24/01/2017 in Macerata, Perugia, Rieti, Ascoli Piceno, L'Aquila, Teramo, Terni and Fermo
     Provinces (comprised the respective coasts) with magnitudes not smaller than 3.1 is presented. \newline
(c) Comparison among $\Gamma^{(i)}_{ZML}$,
    $\Gamma^{(ii)}_{ZML}$ and $\Gamma^{(iii)}_{ZML}$ as $\bar{z}$ varies. The case of
    earthquakes registered from 16/04/2005 to 31/03/2017 in Italy with magnitudes not smaller than 2.5 is presented.}
    \label{GammaZML_compared}
\end{figure}
\end{center}

\begin{center}
\begin{figure}
    \centering
    \includegraphics[scale = 0.6]{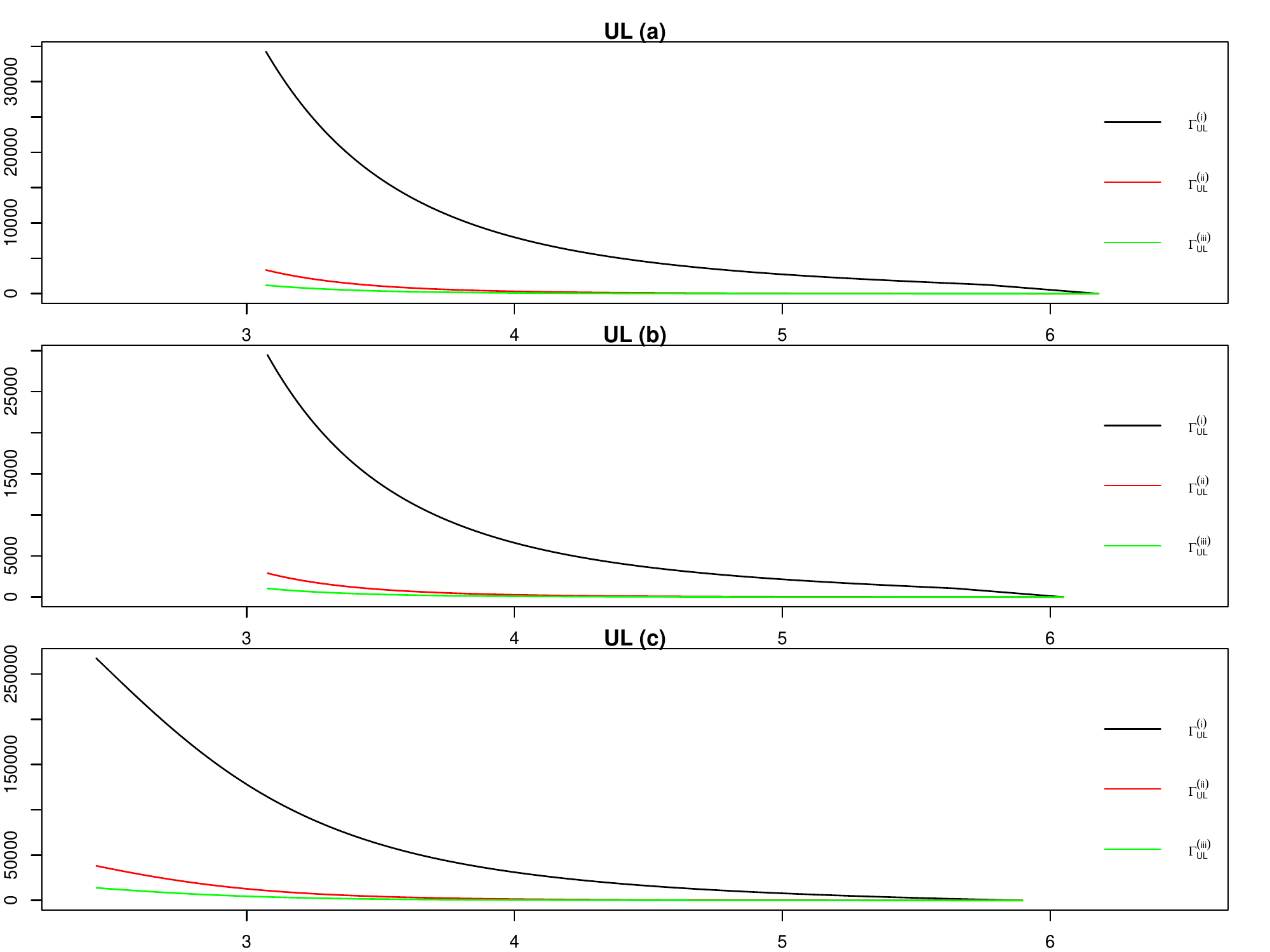}\\
    \caption{(a) Comparison among $\Gamma^{(i)}_{UL}$,
    $\Gamma^{(ii)}_{UL}$ and $\Gamma^{(iii)}_{UL}$ as $\bar{z}$ varies. The case of earthquakes
    registered from 24/01/2016 to 24/01/2017 in Italy with magnitudes not smaller than 3.1 is presented. \newline
    (b) Comparison among $\Gamma^{(i)}_{UL}$,
    $\Gamma^{(ii)}_{UL}$ and $\Gamma^{(iii)}_{UL}$ as $\bar{z}$ varies. The case of earthquakes registered
     from 24/01/2016 to 24/01/2017 in Macerata, Perugia, Rieti, Ascoli Piceno, L'Aquila, Teramo, Terni and Fermo
      Provinces (comprised the respective coasts) with magnitudes not smaller than 3.1 is presented. \newline
(c) Comparison among $\Gamma^{(i)}_{UL}$,
    $\Gamma^{(ii)}_{UL}$ and $\Gamma^{(iii)}_{UL}$ as $\bar{z}$ varies. The case of earthquakes
     registered from 16/04/2005 to 31/03/2017 in Italy with magnitudes not smaller than 2.5 is presented.}
    \label{GammaUL_compared}
\end{figure}
\end{center}

The other cases of cost indicators $\Gamma$s are properly estimated
through standard numerical techniques. Specifically, the generic
interval $[0,\bar{r}]$ is discretized in $S$ sub--intervals with a
discretization step $\Delta r$, so that
$$r_0=0, \qquad r_{s}=r_{s-1}+\Delta r, \qquad r_S=\bar{r}. $$
From such a discretization, the generic integrals defining the
$\Gamma$'s are approximated as follows:
$$
\Gamma=\int_0^{\bar{r}}H(r)dr \sim \Delta r \cdot \sum_{s=1}^S
H(r_s).
$$
Now, recall that a specific value of $\bar{r}$ is associated to a
value of $\bar{z}$. Thus, we can compare the cost indicators in
terms of the threshold magnitudes $\bar{z}$.
\newline
Figure \ref{cost_analysis} allows the comparison among the cases of
$\Gamma_{ZML}$'s and $\Gamma_{UL}$'s as $\bar{z}$ varies,
respectively. The discretization step used for integral
approximation in (\ref{Gamma^iiZML}), (\ref{Gamma^iiiZML}) and
(\ref{Gamma^iiiUL}) is
taken as $\Delta r=$ 0.01.\\
Cost indicators are decreasing functions of $\bar{z}$, as expected.
The value of $\bar{z}$ that represents a measure of the Italian
infrastructures' ability of resisting to earthquakes. 
\newline
The costs decays have no differences in the behaviours considering
the two fit functions (see Figure \ref{cost_analysis}).
\newline
As expected, for both of cases of Eq. (\ref{ZML}) and (\ref{UL}),
the most expensive case emerges by transforming magnitudes into cost
with the exponential function $\Gamma^{(i)}$, while the logarithmic
transformation of the magnitudes leads to the lowest level of cost
indicator and the sensitiveness to increments of $\bar{z}$ are less
evident.
\newline
The $\Gamma^{(ii)}$'s and $\Gamma^{(iii)}$'s decay quite
simultaneously, even if starting by different point, and converge to
zero, while $\Gamma^{(i)}_{ZML}$ and $\Gamma^{(i)}_{UL}$ tend to
rapidly reduce the cost until $\bar{z}$ is around 3.7 (by a visual
inspection). After this threshold the curves' inclination decrease
very slowly denoting resistance to damages reduction.
\newline
Furthermore, the exponential transformations of estimated magnitude
flatten after about $\bar{z} = 3.5$.
\newline
Moreover, one can observe a change in the concavity of the curves
$\Gamma^{(i)}$'s around magnitude 5.7. After such a value, the
curves decrease rapidly to zero. This finding suggests that the
aggregated economic costs of the earthquakes collapse rapidly above
a large enough threshold, and this should be viewed as a hint to the
policymakers of implementing strategies for letting the no-damage
zone above such a magnitude threshold.
\newline
In order to visualize the robustness of the results obtained with
this cost analysis, in Figures \ref{GammaZML_compared} and
\ref{GammaUL_compared} we also present the different curves obtained
from the different dataset presented in Section \ref{robustness}.
Panel (a) is the case of the original sample, (b) is the local
analysis and (c) is the global one.
\newline
For the cases of the cost indicators calibrated on the Eq.
(\ref{ZML}), see Figure \ref{GammaZML_compared}. We can note that
(a) and (b) have the same shapes, but (b) is a little bit scaled due
to the fact that the zones individuated entails the exclusion of
some seismic events. The decays are the same but the curves of the
(b) case reach zero first. A motivation can be found in the
exclusion of an important earthquake of magnitude around 5.5 in the
local dataset, hence leading to slightly cheaper damages. Case (c)
is referred to a wider time window (about 12 years) and to a dataset
with $M_c=2.5$ on average. Consequently, as expected, the increased
amount of minor earthquakes rises the cost mainly in the left side
of the curve. In this case, null costs are achieved at magnitudes
around 5.5. This misrepresentation is due to the functional form of
ZML, being the percentage of high-magnitudes phenomena over the
considered series very low.
\newline
The costs analysis performed with the employment of Eq. (\ref{UL})
are reported in Figure \ref{GammaUL_compared}. For cases (a) and
(b), the same arguments carried out above can be applied. The null
costs are achieved for a magnitude in case (b) smaller than that of
case (a), due to the removal of one important seismic event in the
local dataset. The (c) case is different. There one can appreciate
the relevant capacity of the UL in representing the data. In fact
the zeroing of the costs occurs near magnitude 6.5, which is the
real value of the highest registered earthquake.
\newline
To conclude, the definition of economic costs performed over the
original sample (see Figures \ref{GammaZML_compared} and
\ref{GammaUL_compared}, panel (a)) can be reasonably considered
valid because they coherently represent the logic of the phenomena
that we are studying. Furthermore, the implemented selection of the
local dataset does not change the substance of the findings, hence
supporting the negligibility of space effects in the considered
sample (see De Natale et al. (1988)). Furthermore, results are
robust also in terms of the catalog incompleteness problem, in that
taking magnitudes not smaller than 3.1 and 2.5 has a very weak
effect on the total cost aggregation. 

\section{Conclusions}

This paper deals with a rank-size analysis of earthquakes'
magnitudes occurred in Italy from  $24^{th}$ January, 2016 to
$24^{th}$ January, 2017. Two different fit functions are proposed:
the ZML (see Eq \ref{ZML}) and the UL (see Eq. \ref{UL}). It is
shown that the
earthquakes data exhibit a strong rank-size regularity and that the both functions exhibit a remarkable goodness of fit.\\
The five parameters UL (\ref{UL}) improves the fit -- even if in a
not so significant way -- only when an enlargement in time and
magnitude of the dataset is implemented. In this case, UL is more
capable than ZML to capture the effect of higher earthquakes.
\newline
To e consistent under a seismological perspective, both problems of
incomplete catalog and of space effects have been treated.
\newline
Moreover, a new formulation of economic cost indicators has been
introduced. Such a conceptualization moves from the presence of a
critical threshold for the magnitude which distinguishes earthquakes
in terms of damages.
\newline
The definition of economic costs performed over the original sample
(see Figures \ref{GammaZML_compared} and \ref{GammaUL_compared},
panel (a)) can be reasonably considered valid because they
coherently represent the logic of the phenomena that we are
studying. Furthermore, the implemented selection of the local
dataset does not change the substance of the findings, hence
supporting the negligibility of space effects in the considered
sample (see De Natale et al. (1988)). Results are robust also in
terms of the catalog incompleteness problem, in that taking
magnitudes not smaller than 3.1 and 2.5 has a very weak effect on
the total cost aggregation.
\newline
The analysis of the cost indicators explains clearly that the
reduction of the earthquakes' impact on infrastructures should be
pursue by letting the no-damages magnitude growing (see Figures
\ref{cost_analysis}, \ref{GammaZML_compared} and
\ref{GammaUL_compared}). More than this, the discussion of three
different scenarios for the individual cost of an earthquake with a
given magnitude illustrates also the way in which such a reduction
takes place. The obtained results suggest to adopt risk management
strategies pointing at the mechanism of economic costs creation in
terms of earthquake magnitudes.

\end{document}